\documentclass[%
 aip,
 amsmath,amssymb,
 reprint,%
]{revtex4-1}

\usepackage{graphicx}
\usepackage{dcolumn}
\usepackage{bm}

\usepackage[utf8]{inputenc}
\usepackage[T1]{fontenc}
\usepackage{mathptmx}
\usepackage{etoolbox}

\usepackage{ulem}
\usepackage{xcolor}

\makeatletter
\def\@email#1#2{%
 \endgroup
 \patchcmd{\titleblock@produce}
  {\frontmatter@RRAPformat}
  {\frontmatter@RRAPformat{\produce@RRAP{*#1\href{mailto:#2}{#2}}}\frontmatter@RRAPformat}
  {}{}
}%
\makeatother
\begin{document}

\preprint{AIP/123-QED}

\title{Energy stability analysis of turbulent incompressible flow based on the triple decomposition of the velocity gradient tensor}
\author{J. Hoffman}
 \email{jhoffman@kth.se}
 \homepage{http://www.kth.se/profile/~jhoffman}
\affiliation{KTH Royal Institute of Technology, 10044 Stockholm, Sweden}%

\date{\today}

\begin{abstract}
In the context of flow visualization
a triple decomposition of the velocity gradient
into irrotational straining flow, shear flow and rigid body rotational flow
was proposed by Kol\'a{\v r} in 2007 
[V. Kol\'a{\v r}, International journal of heat and fluid flow, {\bf 28}, 638, (2007)], which has recently  received renewed interest.
The triple decomposition opens for a refined energy stability analysis of the Navier-Stokes equations, with implications for the mathematical analysis of the structure, computability and regularity of turbulent flow. We here perform an energy stability analysis of turbulent incompressible flow, which suggests a scenario where 
at macroscopic scales
any exponentially unstable irrotational 
straining
flow structures rapidly evolve 
towards linearly unstable shear flow and stable rigid body rotational flow. This scenario does not rule out irrotational straining flow close to the Kolmogorov microscales, 
since there viscous dissipation stabilizes the unstable flow structures.  
In contrast to worst case energy stability estimates, this refined stability analysis reflects the existence of stable flow structures in turbulence over extended time.
\end{abstract}

\maketitle

%


By the Navier-Stokes equations, turbulent incompressible flow can be described in terms of its velocity and pressure fields. 
The velocity gradient tensor expresses the local spatial change of the velocity field and can be used, for example, to identify and visualize vortex structures.
But the velocity gradient tensor also plays a key role in a flow stability analysis.

In a domain $\Omega\subset \mathbb{R}^3$ over a time interval $[0,T]$, incompressible flow with constant density $\rho >0$ 
is modeled by the Navier-Stokes equations, 
\begin{eqnarray*}
\rho(\dot u + (u\cdot \nabla )u) - \nabla \cdot \sigma &=& \rho f,\\ 
\nabla \cdot u &=& 0,\\
u(x,0) &=& u_0(x),
\end{eqnarray*}
for $(x,t) \in \Omega\times [0,T]$, together with boundary conditions. 
Here 
$u=u(x,t)=(u_1(x,t),u_2(x,t),u_3(x,t))$ 
is the velocity vector, and $\rho f=\rho f(x,t)$ is the external force vector. 
For a Newtonian fluid, the Cauchy stress tensor $\sigma $ is defined by 
$
\sigma = 2\mu S(u) - \bar pI,
$ 
with $\bar p=\bar p(x,t)$ the mechanical pressure, $\mu$ the dynamic viscosity, and 
\begin{equation}
S(u) = \frac{1}{2}(\nabla u+\nabla u^T)
\label{strain_tensor}
\end{equation}
is the strain rate tensor. 
Since $\nabla \cdot u=0$, it follows that 
$$
\nabla \cdot \sigma = \nabla \cdot (2\mu S(u) - \bar p I) = \mu \Delta u - \nabla \bar p . 
$$
With the kinematic viscosity $\nu=\mu/\rho$, and the scaled pressure $p=\bar p/\rho$, we write the Navier-Stokes equations as 
\begin{eqnarray}
\dot u + (u\cdot \nabla )u + \nabla p -\nu \Delta u &=& f, \label{nse1}\\ 
\nabla \cdot u &=& 0, \label{nse2}\\
u(x,0) &=& u_0(x). \label{nse3}
\end{eqnarray}

If $(u,p)$ is sufficiently regular to satisfy equations (\ref{nse1})-(\ref{nse2}) for each $(x,t)\in \Omega \times [0,T]$, then it is referred to as a classical solution, or strong solution.
Hence, with $(u,p)$ a classical solution we can take the scalar product of equation (\ref{nse1}) with $u$ 
and multiply equation (\ref{nse2}) by $p$, then add them together and integrate over the domain $\Omega$. 
By the divergence theorem we obtain the well known global energy balance
\begin{equation}
 \frac{d}{dt}
\int_{\Omega} \frac{1}{2} \vert u\vert^2\, dx = 
\int_{\Omega} f\cdot u\, dx 
- \nu \int_{\Omega} \vert \nabla u\vert ^2 \, dx . 
\label{energy_eq}
\end{equation}

We here neglect the contribution from the
boundary $\partial \Omega$, 
using a standard assumption that the solution decays towards the boundary. 
If we include boundary effects in the analysis, additional boundary integrals appear.
The energy balance (\ref{energy_eq}) expresses the injection of kinetic energy by external forces and loss of kinetic energy by viscous dissipation. 
The energy dissipation can equivalently be expressed as
\begin{equation}
    \nu \int_{\Omega} \vert \nabla u\vert ^2 \, dx = 
    \nu \int_{\Omega} \vert \omega\vert ^2 \, dx = 
    \nu \int_{\Omega} 2\vert \Omega\vert ^2 \, dx, 
\end{equation}
where $\omega=\nabla \times u$ is the vorticity 
vector. 
The spin tensor
\begin{equation}
    \Omega(u) = \frac{1}{2}(\nabla u - \nabla u^T), 
\end{equation}
together with the strain rate tensor (\ref{strain_tensor}) constitute the standard 
double
decomposition of the velocity gradient, 
\begin{equation}
    \nabla u = S(u) + \Omega(u).
    \label{double_decomp}
\end{equation}


The evolution of vorticity is governed by the equation 
\begin{eqnarray}
\dot \omega + (u \cdot \nabla ) \omega - (\omega \cdot \nabla ) u - \nu \Delta \omega &=& 0, \label{vort1}\\ 
\omega(x,0) &=& \omega_0(x),  \nonumber
\end{eqnarray}
obtained by taking the curl of the Navier-Stokes equations, and analogous to the balance of kinetic energy (\ref{energy_eq}) 
we can take the scalar product of the equation with the vorticity vector and integrate, to get
\begin{eqnarray}
&& \frac{d}{dt} \int_{\Omega }\frac{1}{2} \vert \omega \vert ^2 \, dx \label{vort_energy}
= 
\int_{\Omega} \omega^T \nabla u\, \omega \, dx 
- \nu \int_{\Omega} \vert \nabla \omega \vert^2 \, dx .
\end{eqnarray}
The dissipation term is 
analogous
in the energy balance (\ref{energy_eq}), but
here
is also a new term with properties determined by the velocity gradient. 
For example, a velocity gradient which reflects a straining 
flow
leads to an exponential growth of vorticity referred to as vortex stretching, 
which
in turbulence 
is 
a mechanism 
that is
responsible for the transfer of energy to smaller scales. 

Vorticity growth is also used 
as a criterion
to detect blow-up of classical solutions to the Navier-Stokes equations \cite{beale1984remarks}. 
Whether or not it always exists a classical solution is one of the Clay Institute \$1 million Prize problems \cite{fefferman2006existence}. 
If blow-up can occur, that would support a mathematical model of turbulence in the form of weak solutions that satisfy the equations only in an average sense \cite{leray1934mouvement}, and which dissipate energy 
independent of viscosity
through the appearances of singularities  \cite{duchon2000inertial}, in line with the Onsager conjecture \cite{onsager1949statistical}.

Very similar in structure to the vorticity equation (\ref{vort1}) is the system of pertubation equations, which describe the evolution of a velocity perturbation $\varphi=\varphi(x,t)$ and a pressure perturbation $\theta=\theta(x,t)$ in a flow field $(u,p)$ governed by the Navier-Stokes equations, 
\begin{eqnarray}
\dot \varphi + (u \cdot \nabla ) \varphi + (\varphi \cdot \nabla ) u + (\varphi \cdot \nabla ) \varphi + \nabla \theta - \nu \Delta \varphi &=& 0, \nonumber \\ 
\nabla \cdot \varphi &=& 0, \label{pert2} \\
\varphi(x,0) &=& \varphi_0(x).  \nonumber 
\end{eqnarray}
In the vorticity equation the pressure term from the Navier-Stokes equations disappears because $\nabla \times \nabla p=0$, and the vorticity field is divergence free since $\nabla \cdot \omega = \nabla \cdot \nabla \times u=0$. 
The perturbation equations are studied in the fields of hydrodynamic stability and flow control, together with the associated adjoint Navier-Stokes equations 
\cite{schmid2002stability}. 

In the context of adaptive numerical methods, the adjoint Navier-Stokes equations are used for goal oriented a posteriori error estimation \cite{hoffman2007computational} and take the following form,  
\begin{eqnarray*}
- \dot \lambda - (u \cdot \nabla ) \lambda + \nabla U^T \lambda - \nabla \pi - \nu \Delta \lambda &=& 0,\\ 
\nabla \cdot \lambda &=& 0,\\
\lambda(x,T) &=& \lambda_T(x), 
\end{eqnarray*}
with $\lambda = \lambda(x,t)$ the adjoint velocity vector, and $\pi =\pi(x,t)$ the scalar adjoint pressure. 
Here $U=U(x,t)$ is a computed approximation of the velocity, and  
$\nabla U^T$ is the transpose of $\nabla U$. 
The goal of the computation is an output of interest which is defined in terms of 
sources
or boundary conditions in the adjoint equations \cite{hoffman2015towards}. 
Note that the adjoint equations evolve backwards in time from a final condition at $t=T$, 
but with the change of variables $s=T-t$ we instead obtain a problem that evolves from $s=0$ to $s=T$, 
\begin{eqnarray}
\dot \lambda - (u \cdot \nabla ) \lambda + \nabla U^T \lambda - \nabla \pi - \nu \Delta \lambda &=& 0, \nonumber \\
\nabla \cdot \lambda &=& 0, \label{anse2} \\
\lambda(x,0) &=& \lambda_T(x). \nonumber 
\end{eqnarray}

The similarities between the perturbation equations, the adjoint equations and the vorticity equation are clear. 
Specifically, in an energy analysis they all share the same stability properties.
From the perturbation equations 
we obtain an energy balance for the perturbations, the Reynolds-Orr energy equation \cite{serrin1959stability}, 
\begin{eqnarray}
&& \frac{1}{2}\frac{d}{dt} \int_{\Omega }\vert \varphi \vert ^2 \, dx 
= 
- \int_{\Omega} \varphi^T \nabla u \, \varphi \, dx 
- \nu \int_{\Omega} \vert \nabla \varphi \vert^2 \, dx , \label{pert_energy}
\end{eqnarray}
and similar for the adjoint equations (\ref{anse2}), 
\begin{eqnarray}
&& \frac{1}{2}\frac{d}{ds} \int_{\Omega }\vert \lambda \vert ^2 \, dx 
= 
- \int_{\Omega} \lambda^T \nabla U^T\, \lambda \, dx 
- \nu \int_{\Omega} \vert \nabla \lambda \vert^2 \, dx . \label{anse_energy}
\end{eqnarray}
%
Recall that $\nabla U\approx \nabla u$, and by the definition of the scalar product we have that 
\begin{equation}
\int_{\Omega} \varphi^T \nabla u \, \varphi \, dx = 
\int_{\Omega} (\nabla u \, \varphi )^T \varphi \, dx = 
\int_{\Omega} \varphi^T \nabla u^T \varphi \, dx. \label{scalar_prod}
\end{equation}
Therefore, the key to any stability analysis is the term 
\begin{equation}
P(u) = \pm \int_{\Omega} \phi^T \nabla u\, \phi \, dx, \quad \phi=\varphi, \lambda, \omega. 
\label{energy_prod}
\end{equation}

To analyze $P(u)$, a natural starting point is the classical decomposition (\ref{double_decomp}) into the symmetric strain rate tensor $S(u)$ and
the skew-symmetric spin tensor $\Omega(u)$. 
By equation (\ref{double_decomp}) and equation (\ref{scalar_prod}), it follows that 
\begin{equation}
\int_{\Omega} \phi^T \nabla u \, \phi \, dx = 
\int_{\Omega} \phi^T S(u) \, \phi \, dx. 
\label{stab_strain}
\end{equation}
That is, only $S(u)$ contributes to the critical term (\ref{energy_prod}). 


To better understand the stability of different flow structures, we recall three 
simple
archetypical solutions to the Navier-Stokes equations.  
Let 
$(u_{rr},p_{rr})$
denote rigid body rotational flow about the $x_1$ coordinate axis, 
\begin{equation}
u_{rr}(x)=(0,-x_3,x_2),\quad p_{rr}=\frac{1}{2}(x_2^2+x_3^2),    
\label{arch_rot}
\end{equation} 
an example for which $S(u_{rr})=0$ but $\Omega(u_{rr}) \neq 0$. 
In contrast,  irrotational straining flow (elongation 
and compression) in the $x_1x_2$ plane 
is an example for which $\Omega(u_{el})=0$ but $S(u_{el}) \neq 0$, 
\begin{equation}
u_{el}(x)=(x_1,-x_2,0), \quad p_{el}=-\frac{1}{2}(x_1^2+x_2^2).  
\label{arch_strain}
\end{equation} 
This follows from inspection of the velocity gradients, 
$$
\nabla u_{rr} = 
\left[\begin{array}{ccc}
0 & 0 & 0 \\
0& 0 & -1 \\
0& 1& 0
\end{array}\right],
\quad 
\nabla u_{el} = 
\left[\begin{array}{ccc}
1 & 0 & 0 \\
0& -1 & 0 \\
0& 0& 0
\end{array}\right]. 
$$

One might then be lead to believe that equation (\ref{double_decomp}) decomposes the flow in every point 
of the domain into a sum of rigid body rotational motion and irrotational straining motion.
This, however, does not take shear flow into consideration. An example of shear flow in the $x_2x_3$ plane is  
\begin{equation}
u_{sh}(x)=(0,-x_3,0), \quad p=0, 
\label{arch_shear}
\end{equation} 
with gradient 
$$
\nabla u_{sh} = 
\left[\begin{array}{ccc}
0 & 0 & 0 \\
0& 0 & -1 \\
0& 0& 0
\end{array}\right], 
$$
which contributes to both the strain rate tensor and the spin tensor,  
$$
S(u_{sh}) = 
\frac{1}{2}
\left[\begin{array}{ccc}
0 & 0 & 0 \\
0& 0 & -1 \\
0& -1& 0
\end{array}\right], 
\quad 
\Omega(u_{sh}) = 
\frac{1}{2}
\left[\begin{array}{ccc}
0 & 0 & 0 \\
0& 0 & -1 \\
0& 1& 0
\end{array}\right]. 
$$

Hence, the decomposition (\ref{double_decomp}) is not able to distinguish between the three archetypes, each with its own unique properties. From equation (\ref{stab_strain}) we conclude that $u_{rr}$ does not contribute to the critical term (\ref{energy_prod}), therefore, $u_{rr}$ is stable in the sense that the only active terms in the energy balances (\ref{vort_energy}), (\ref{pert_energy}) and (\ref{anse_energy}) are the dissipative terms. 
The critical term (\ref{energy_prod}) for the three examples takes the forms 
\begin{eqnarray}
P(u_{rr}) &=& 0,\\ 
P(u_{el}) &=& \pm \int_{\Omega} (\vert \phi_1\vert ^2 - \vert \phi_2\vert ^2)\, dx, \\
P(u_{sh}) &=& \mp \int_{\Omega} \phi_2 \phi_3 \, dx. 
\end{eqnarray}

To determine the stability of $u_{el}$ and $u_{sh}$, it is instructive to analyze the energy in each vector component individually. %
In the case of the vorticity, we take the scalar product of equation (\ref{vort1}) with each of the vectors
\begin{equation*}
    (\omega_1,0,0), \quad (0,\omega_2,0), \quad (0,0,\omega_3), 
\end{equation*}
and then integrate over the domain $\Omega$, 
to get the energy balance for each component of the vorticity vector. For the irrotational straining flow (\ref{arch_strain}) this leads to 
\begin{eqnarray*}
\frac{1}{2}\frac{d}{dt} \int_{\Omega}\vert \omega_1\vert ^2 \, dx &=& \int_{\Omega} \vert \omega_1\vert ^2 \, dx - \nu \int_{\Omega} \vert \nabla \omega_1\vert^2\, dx, \\
\frac{1}{2}\frac{d}{dt} \int_{\Omega}\vert \omega_2\vert ^2 \, dx &=& - \int_{\Omega} \vert \omega_2\vert ^2 \, dx - \nu \int_{\Omega} \vert \nabla \omega_2\vert^2\, dx, \\
\frac{1}{2}\frac{d}{dt} \int_{\Omega}\vert \omega_3\vert ^2 \, dx &=& - \nu \int_{\Omega} \vert \nabla \omega_3\vert^2\, dx, 
\end{eqnarray*}
from which we detect exponential growth of the energy in $\omega_1$, if it is not damped by the viscous dissipation. The growth is aligned with the direction of elongation in $u_{el}$, reflecting the well known vortex stretching mechanism. 
Note that $\omega_1$ may represent either rigid body rotational flow or shear flow, or a combination thereof. 

A similar componentwise analysis of the perturbation equations (\ref{pert2}) gives, 
\begin{eqnarray*}
\frac{1}{2}\frac{d}{dt} \int_{\Omega}\vert \varphi_1\vert ^2 \, dx &=& - \int_{\Omega} \vert \varphi_1\vert ^2 \, dx - \nu \int_{\Omega} \vert \nabla \varphi_1\vert^2\, dx - \int_{\Omega} \frac{\partial \theta}{\partial x_1} \varphi_1\, dx, \\
\frac{1}{2}\frac{d}{dt} \int_{\Omega}\vert \varphi_2\vert ^2 \, dx &=& \int_{\Omega} \vert \varphi_2\vert ^2 \, dx - \nu \int_{\Omega} \vert \nabla \varphi_2\vert^2\, dx - \int_{\Omega} \frac{\partial \theta}{\partial x_2} \varphi_2\, dx, \\
\frac{1}{2}\frac{d}{dt} \int_{\Omega}\vert \varphi_3\vert ^2 \, dx &=& - \nu \int_{\Omega} \vert \nabla \varphi_3\vert^2\, dx - \int_{\Omega} \frac{\partial \theta}{\partial x_3} \varphi_3\, dx, 
\end{eqnarray*}
with two notable differences from the analysis of the vorticity. First, the critical term (\ref{energy_prod}) has the opposite sign, hence, exponential growth here appears in the second component $\varphi_2$. Second, the individual components of the pressure perturbation gradient appear, even though they cancel when summed up in the Reynolds-Orr equation (\ref{pert_energy}) due to the fact that $\varphi$ is divergence free. 

In the case of the shear flow (\ref{arch_shear}), the analogous stability analysis gives  
\begin{eqnarray*}
\frac{1}{2}\frac{d}{dt} \int_{\Omega}\vert \omega_1\vert ^2 \, dx &=& - \nu \int_{\Omega} \vert \nabla \omega_1\vert^2\, dx, \\
\frac{1}{2}\frac{d}{dt} \int_{\Omega}\vert \omega_2\vert ^2 \, dx &=& - \int_{\Omega} \omega_2\omega_3 \, dx - \nu \int_{\Omega} \vert \nabla \omega_2\vert^2\, dx, \\
\frac{1}{2}\frac{d}{dt} \int_{\Omega}\vert \omega_3\vert ^2 \, dx &=& - \nu \int_{\Omega} \vert \nabla \omega_3\vert^2\, dx, 
\end{eqnarray*}
and 
\begin{eqnarray*}
\frac{1}{2}\frac{d}{dt} \int_{\Omega}\vert \varphi_1\vert ^2 \, dx &=& - \nu \int_{\Omega} \vert \nabla \varphi_1\vert^2\, dx, \\
\frac{1}{2}\frac{d}{dt} \int_{\Omega}\vert \varphi_2\vert ^2 \, dx &=& \int_{\Omega} \varphi_2 \varphi_3 \, dx - \nu \int_{\Omega} \vert \nabla \varphi_2\vert^2\, dx, \\
\frac{1}{2}\frac{d}{dt} \int_{\Omega}\vert \varphi_3\vert ^2 \, dx &=& - \nu \int_{\Omega} \vert \nabla \varphi_3\vert^2\, dx, 
\end{eqnarray*}
from which we observe that no exponential growth takes place, but instead potential linear growth of the energy in $\omega_2$ and $\varphi_2$, with $x_2$ the main flow direction in equation (\ref{arch_shear}).    

We conclude that the irrotational straining flow (\ref{arch_strain}) is exponentially unstable, that the shear flow (\ref{arch_shear}) is linearly unstable, and that the rigid body rotational flow (\ref{arch_rot}) is stable. These archetypical solutions to the Navier-Stokes equations are clearly very simple, but we will find that they in their simplicity hold the essence of the stability properties of a real turbulent flow. 


The discrepancy between the double decomposition (\ref{double_decomp}) and the three archetypical flow structures illustrates a problem of both practical and pedagogical nature. 
The spin tensor does not isolate rigid body rotational flow but may also include
shear flow, which is a well known complication in the context of flow visualization \cite{zhan2019comparison,kolavr2020consequences,beaumard2019importance}. 
For example, in a shear layer the spin tensor is large while 
there may be no rigid rotation what so ever. 
A pedagogical problem arises in the understanding of vortex lines, tubes, filaments and sheets, all defined in terms of vorticity that can mean either shear flow or rigid body rotational flow,
or a combination of the two. 

In 2007, Kol\'a{\v r} \cite{kolavr2007vortex} proposed a triple decomposition of the velocity gradient tensor, 
which corresponds to the three archetypes of rigid body rotational flow, irrotational straining flow and shear flow. 
The key idea is to first use an orthogonal matrix $Q$ and its transpose $Q^T=Q^{-1}$ to transform the velocity gradient tensor into a frame of reference where a shear flow tensor $(\nabla u)_{SH}$ can be isolated, 
\begin{equation}
\nabla u = Q^T((\nabla u)_{RES} + (\nabla u)_{SH})Q, 
\label{shearsubtract}
\end{equation}
after which 
the remaining residual tensor $(\nabla u)_{RES}$ is split into a symmetric tensor $(\nabla u)_{EL}$ and a skew-symmetric tensor $(\nabla u)_{RR}$, 
corresponding to irrotational strain and rigid body rotation, respectively,  
\begin{eqnarray}
(\nabla u)_{RES} &=& 
\frac{1}{2}((\nabla u)_{RES} + (\nabla u)_{RES}^T) + 
\frac{1}{2}((\nabla u)_{RES} - (\nabla u)_{RES}^T) \nonumber \\
&=& 
(\nabla u)_{EL} + (\nabla u)_{RR}. \label{triple_dec0}
\end{eqnarray}
This amounts to the triple decomposition
\begin{equation}
\nabla u = Q^T ((\nabla u)_{EL} + (\nabla u)_{RR} + (\nabla u)_{SH}) Q , 
\label{triple_dec}
\end{equation}
which states that the velocity gradient tensor can be decomposed into a sum of
irrotational strain, rigid body rotation and shear flow. 

Equation (\ref{triple_dec0}) follows the standard double decomposition (\ref{double_decomp}), whereas the decomposition in equation (\ref{shearsubtract}) needs to be defined.  
Kol\'a{\v r} gave the following definition of the residual tensor,  
\begin{equation}
[\nabla u_{RES}]_{ij} = \mbox{sgn}([\overline{\nabla u}]_{ij}) \min([\overline{\nabla u}]_{ij}, [\overline{\nabla u}]_{ji}),
\label{kolarres}
\end{equation}
with $[\cdot]_{ij}$ each tensor component and $\overline{\nabla u} = Q\, \nabla u\, Q^T$, 
where $Q$ should be determined such that the residual tensor is minimized in the Frobenius norm $\Vert \cdot \Vert _F$, 
referred to as the {\it basic reference frame}. 
By the properties of the norm and the orthogonal matrix $Q$, minimization of the residual tensor is equivalent to maximization of the 
quantity $$\Delta(Q) = \vert [\overline{\nabla u}]^2_{12}-[\overline{\nabla u}]^2_{21}\vert  + \vert [\overline{\nabla u}]^2_{13} - [\overline{\nabla u}]^2_{31}\vert  + \vert [\overline{\nabla u}]^2_{23} - [\overline{\nabla u}]^2_{32}\vert,$$
since $\nabla u$ is independent of $Q$ and 
\begin{eqnarray*}
    \Vert \nabla u\Vert_F^2 &=& 
    \Vert Q^T ((\nabla u)_{RES} + (\nabla u)_{SH}) Q\Vert_F^2 \\
    &=& \Vert (\nabla u)_{RES} + (\nabla u)_{SH}\Vert_F^2 \\
    &=& \Vert (\nabla u)_{RES}\Vert_F^2 + \Delta(Q).
\end{eqnarray*}
Kol\'a{\v r} first solved this optimization problem for two dimensional flow \cite{kolavr2007vortex} and later also for three dimensional flow \cite{kolavr2011triple}. More recently, new algorithms have been proposed to solve the same optimization problem \cite{nagata2020triple,boukharfane2021triple}.

Another strategy is to identify the shear flow tensor with the non-normal part of the velocity gradient tensor, in which case equation (\ref{shearsubtract}) follows from the Schur decomposition of the velocity gradient tensor. The Schur decomposition of a general (possibly complex) $3\times 3$ matrix $A$ takes the form 
\begin{equation}
A = U^* T U = U^*(D + R)U,
\label{schur}
\end{equation}
with $U$ a unitary matrix with adjoint $U^*=U^{-1}$, $T$ an upper triangular matrix with the eigenvalues of $A$ on its diagonal, $D$ the diagonal part of $T$, and $R = T-D$ its strictly upper triangular part. The eigenvalues $\lambda_1,\lambda_2,\lambda_3$ are unique, but how they are ordered on the diagonal of $T$ is not. $U$ is not unique, and neither is $R$ although the Frobenius norm of $R$ is, since by equation (\ref{schur}), 
\begin{equation}
    \Vert R\Vert _F^2  = \Vert A\Vert_F^2 - \Vert D\Vert_F^2  
    = \sum_{i=1}^3 \sigma_i^2 - \sum_{i=1}^3 \vert \lambda_i\vert ^2, 
    \label{schurfrob}
\end{equation}
with $\sigma_1,\sigma_2,\sigma_3$ the singular values of $A$. 
If $A$ is a normal matrix then $R=0$. Hence, the matrix $U^*DU$ represents the normal part of $A$, and $U^*RU$ the non-normal part. 
By identifying the non-normal part of the velocity gradient tensor with the shear flow tensor $(\nabla u)_{SH}=R$, the Schur decomposition can be used to construct the triple decomposition \cite{keylock2018schur}.

$(\nabla u)_{SH}=R$ 
also satisfies equation (\ref{kolarres}), 
possibly using complex arithmetic if the velocity gradient tensor has complex eigenvalues, corresponding to its skew-symmetric part being non-zero $(\nabla u)_{RR}\neq 0$. 
Complex eigenvalues of a real matrix always exist as pairs of complex conjugates $(\lambda,\bar \lambda)=(a+ib,a-ib)$. Complex arithmetic can be avoided by using the the real Schur decomposition  \cite{datta1995numerical} of the velocity gradient tensor, where each pair of complex eigenvalues are represented by the real  
$2\times 2$ block  
\begin{equation}
    \left[
    \begin{array}{cc}
        a & b \\
        -b & a
    \end{array}
    \right]
    \label{realSchurblock}
\end{equation}
on the diagonal of the matrix $D$. 
Equation (27) is satisfied for the triple decomposition based on the real Schur decomposition, 
since the block (\ref{realSchurblock}) is symmetric in absolute value. 
If $(\nabla u)_{RR} = 0$ the complex and real Schur decompositions are equivalent.  
The Schur decomposition, real or complex, can be constructed e.g. by the QR algorithm, which, therefore, offers an alternative to methods based on the optimization problem 
\cite{keylock2018schur,boukharfane2021compressibility}. 
But the relation between the Schur transformations and the basic reference frame is still an open question.

The non-uniqueness of the Schur decomposition is not as serious as one may first think. 
If $(\nabla u)_{RR}\neq 0$ the flow exhibits rigid body rotation and the velocity gradient tensor has a pair of complex conjugate eigenvalues, and an eigenvector of the remaining real eigenvalue defines the axis of rotation. Hence, the non-uniqueness of the frame of reference is merely a matter of choosing a coordinate system aligned with the axis of rotation \cite{liu2018rortex}. In the case $(\nabla u)_{RR} = 0$ and the velocity gradient tensor is normal, the frame of reference is given by eigenvectors of the three real eigenvalues corresponding to $(\nabla u)_{EL}$.  


In light of the triple decomposition (\ref{triple_dec}), the 
critical term (\ref{energy_prod}) can be decomposed as follows, 
\begin{eqnarray}
P(u) &=& \pm \int_{\Omega} \phi^T Q^T ((\nabla u)_{EL} + (\nabla u)_{RR} + (\nabla u)_{SH}) Q \phi \, dx \nonumber \\
&=& \pm \int_{\Omega} (Q \phi)^T (\nabla u)_{EL} (Q \phi) \, dx  \pm \int_{\Omega} (Q \phi)^T (\nabla u)_{SH} (Q \phi) \, dx \nonumber \\
&=& P_{EL}(u) + P_{SH}(u). \label{energy_prod_triple}
\end{eqnarray}
Both 
$(\nabla u)_{RR}$
and 
$(\nabla u)_{SH}$
have zero diagonal components, and 
$(\nabla u)_{EL}$ 
is a symmetric tensor with zero trace, which follows from the divergence 
free condition of incompressible flow, the cyclic property of the trace operation, and the definition of 
an orthogonal matrix, 
\begin{eqnarray*}
0 = \nabla \cdot u &=& \mbox{tr}(\nabla u) = \mbox{tr}(Q^T\, \overline{\nabla u}\, Q) \\ 
&=& \mbox{tr}(QQ^T \, \overline{\nabla u}) = \mbox{tr}(\overline{\nabla u}) = \mbox{tr}((\nabla u)_{EL}).  
\end{eqnarray*}

Hence, the eigenvalues of the matrix $(\nabla u)_{EL}$ are real and sum to zero, which means that 
either all eigenvalues are zero, or 
there will always be at least one eigenvalue that will generate 
exponential 
growth. 
With the Schur decomposition, $(\nabla u)_{SH}$ 
is a 
non-normal
strictly triangular matrix, which generates linear
growth.
The tensor 
$(\nabla u)_{RR}$
is skew-symmetric with purely imaginary eigenvalues, which corresponds to 
a stable vortex. 

By Taylor's formula and equation (\ref{triple_dec}), near each interior point $x_0\in \Omega$ we can construct a linear approximation of the velocity field, 
\begin{eqnarray}
    u(x) &\approx& u(x_0) + \nabla u(x_0) (x-x_0) \nonumber \\
    &=& u_C + u_{EL}(x) + u_{RR}(x) + u_{SH}(x),\label{vel_decom}
\end{eqnarray}
a decomposition of the local velocity field defined by 
\begin{eqnarray}
u_C &=& u(x_0), \\
u_{EL}(x) &=& (Q^T(\nabla u)_{EL}Q)(x_0)(x-x_0), \\
u_{RR}(x) &=& (Q^T(\nabla u)_{RR}Q)(x_0)(x-x_0), \\
u_{SH}(x) &=& (Q^T(\nabla u)_{SH}Q)(x_0)(x-x_0),
\end{eqnarray}
where each part represents constant flow, irrotational straining flow, rigid body rotational flow and shear flow, respectively. 
Each part has its own specific stability properties, analogous to the archetypes
(\ref{arch_strain}), (\ref{arch_rot}) and (\ref{arch_shear}).

In the energy balances (\ref{vort_energy}), (\ref{pert_energy}) and (\ref{anse_energy}), 
the growth
generated by strain and shear may be damped by viscous dissipation, but 
at macroscopic turbulent scales
the critical term 
(\ref{energy_prod_triple}) 
will dominate. 
Hence, the stability analysis suggests that at macroscopic turbulent scales, 
rigid body rotational flow can exist for a long time, shear flow for a short time, and 
irrotational straining flow 
can only appear
as intermittent very short-lived events.
The decomposition (\ref{vel_decom}) always exists, 
assuming the velocity field is regular enough,
and therefore implies that the flow evolves from exponentially unstable irrotational straining flow and linearly unstable shear flow, towards stable rigid body rotational 
flow.
Mechanisms for this evolution include vortex stretching by strain, and shear layer roll up by the Kelvin-Helmholtz instability \cite{nagata2020triple,watanabe2020characteristics,hayashi2021relation}. 

At microscopic turbulent scales, close to the Kolmogorov scale, viscous dissipation is significant. A detailed understanding of how energy is dissipated in turbulent flow is still lacking, but the stability analysis suggests that irrotational straining flow can be sustained longer at the microscopic scales, 
since it is stabilized by viscous dissipation.
Experimental data also supports that flow structures at microscopic turbulent scales involve strain, for example, elongation of vortices (vortex stretching) and compression of shear layers (strain self-amplification) \cite{moffatt2021extreme,debue2021three,johnson2020energy,nagata2020triple,watanabe2020characteristics}.


The triple decomposition of the velocity gradient tensor (\ref{triple_dec}) together with the linearization (\ref{vel_decom}) gives a characterization of the local velocity field as a sum of a constant part, a strain part, a rigid body rotational part and a shear part. In the energy stability analysis the critical term (\ref{energy_prod_triple}) exhibits the stability property of each part, where strain corresponds to exponential growth and shear to linear growth. At macroscopic turbulent scales viscous dissipation can be neglected, which suggests that straining flow cannot exist other than for brief instants and shear flow only for a short time. Rigid body rotation, on the other hand, is a stable flow structure. In contrast, at microscopic turbulent scales viscous dissipation permits flow structures that include strain and shear to exist over extended time.  

Experimental observations appear to support the predictions of the stability analysis that strain and shear flow are short-lived at macroscopic turbulent scales, to instead evolve into vortex structures through vortex stretching and shear layer roll up \cite{nagata2020triple,watanabe2020characteristics,hayashi2021relation}. Close to the Kolmogorov scale, experimental data recently reported identifies intense vortex stretching \cite{moffatt2021extreme,debue2021three}. 
A detailed characterization of the finest turbulent flow structures could give clues to how energy is dissipated 
\cite{hayashi2021characteristics,keylock2019turbulence,das2020revisiting,nguyen2020characterizing}, 
and help to solve the outstanding mathematical questions of  Onsager's conjecture \cite{onsager1949statistical} and the blow-up problem \cite{fefferman2006existence}. 

The focus of this note is turbulent structures away from boundaries, but the decomposition (\ref{vel_decom}) is valid also near boundaries. It is well know that as the Reynolds number increases, shear flow in laminar boundary layers 
transition
into turbulent boundary layers populated by vortex structures. 

In a large eddy simulation, the turbulent scales are truncated by the resolution of the computational mesh, typically in the inertial range far from the Kolmogorov scale. Hence, the stability analysis predicts that the resolved turbulent flow will be dominated by
a collection of stable rigid body rotational flow structures, vortex tubes, and linearly unstable shear flow structures, shear pancakes, which either dissipate or roll up to form vortex tubes. Strain is short-lived and only appears intermittently in scenarios where the other two types of flow structures are intensified. 
Computability
of mean quantities in turbulent flow can be formulated in terms of a posteriori error analysis and the adjoint equations \cite{hoffman2006new,hoffman2004computability}. In the context of adaptive finite element methods, stable solutions to the adjoint equations have been computed for a wide variety of examples of fully developed turbulent flow 
\cite{hoffman2015towards,de2016computation,hoffman2014time,hoffman2006adaptive,hoffman2005computation,hoffman2009efficient}. 

Based on the stability analysis in this note we conjecture that turbulence is a 
less chaotic phenomenon than is commonly believed, 
in the sense that its elements are robust and predictable, vortex tubes and shear pancakes at macroscopic scales which are intensified by strain to be dissipated at microscopic scales. 
Therefore, it is not surprising that mean quantities that do not depend on the detailed dynamics of these turbulent flow structures, such as the drag of a car, can be predicted in computer simulations, which is also reflected in the adjoint equations. 



Apart from this general characterization of turbulent flow, the triple decomposition can be used in an energy stability analysis of model solutions to the Navier-Stokes equations, such as a Burgers vortex \cite{burgers1948mathematical} which exhibits the stability properties of combined strain, rigid rotation, shear flow and viscous dissipation \cite{schmid2004three}.

\medskip 

\section{Acknowledgements}

The following article has been accepted for publication in Physics of Fluids.
This work is supported by the Swedish Research Council 
(contract number 2018-04854). The author would like to thank Dr. V\'aclav Kol\'a{\v r} for a helpful discussion regarding the optimization problem definition of the triple decomposition.

\section{References}

\nocite{*}
\bibliography{Article_File_letter}

\end{document}